\documentstyle[12pt, bezier]{article}
\def\beq{\begin{equation}}
\def\eeq{\end{equation}}

\def\ap#1#2#3 {Ann. Phys. (NY) {\bf#1} (19#2) #3}
\def\apj#1#2#3 {Astrophys. J. {\bf#1} (19#2) #3}
\def\apjl#1#2#3 {Astrophys. J. Lett. {\bf#1} (19#2) #3}
\def\app#1#2#3 {Acta. Phys. Pol. {\bf#1} (19#2) #3}
\def\ar#1#2#3 {Ann. Rev. Nucl. Part. Sci. {\bf#1} (19#2) #3}
\def\cpc#1#2#3 {Computer Phys. Comm. {\bf#1} (19#2) #3}
\def\err#1#2#3 {{\it Erratum} {\bf#1} (19#2) #3}
\def\ib#1#2#3 {{\it ibid.} {\bf#1} (19#2) #3}
\def\jmp#1#2#3 {J. Math. Phys. {\bf#1} (19#2) #3}
\def\ijmp#1#2#3 {Int. J. Mod. Phys. {\bf#1} (19#2) #3}
\def\jetp#1#2#3 {JETP Lett. {\bf#1} (19#2) #3}
\def\jpg#1#2#3 {J. Phys. G. {\bf#1} (19#2) #3}
\def\mpl#1#2#3 {Mod. Phys. Lett. {\bf#1} (19#2) #3}
\def\nat#1#2#3 {Nature (London) {\bf#1} (19#2) #3}
\def\nc#1#2#3 {Nuovo Cim. {\bf#1} (19#2) #3}
\def\nim#1#2#3 {Nucl. Instr. Meth. {\bf#1} (19#2) #3}
\def\np#1#2#3 {Nucl. Phys. {\bf#1} (19#2) #3}
\def\pcps#1#2#3 {Proc. Cam. Phil. Soc. {\bf#1} (#2) #3}
\def\pl#1#2#3 {Phys. Lett. {\bf#1} (19#2) #3}
\def\prep#1#2#3 {Phys. Rep. {\bf#1} (19#2) #3}
\def\prev#1#2#3 {Phys. Rev. {\bf#1} (19#2) #3}
\def\prl#1#2#3 {Phys. Rev. Lett. {\bf#1} (19#2) #3}
\def\prs#1#2#3 {Proc. Roy. Soc. {\bf#1} (19#2) #3}
\def\ptp#1#2#3 {Prog. Th. Phys. {\bf#1} (19#2) #3}
\def\ps#1#2#3 {Physica Scripta {\bf#1} (19#2) #3}
\def\rmp#1#2#3 {Rev. Mod. Phys. {\bf#1} (19#2) #3}
\def\rpp#1#2#3 {Rep. Prog. Phys. {\bf#1} (19#2) #3}
\def\sjnp#1#2#3 {Sov. J. Nucl. Phys. {\bf#1} (19#2) #3}
\def\spj#1#2#3 {Sov. Phys. JETP {\bf#1} (19#2) #3}
\def\spu#1#2#3 {Sov. Phys. Usp. {\bf#1} (19#2) #3}
\def\zp#1#2#3 {Zeit. Phys. {\bf#1} (19#2) #3}

\begin{document}
\begin{titlepage}
\begin{center}
{\Large \bf Theoretical Physics Institute \\
University of Minnesota \\}  \end{center}
\vspace{0.3in}
\begin{flushright}
TPI-MINN-00/38 \\
UMN-TH-1915-00 \\
July 2000
\end{flushright}
\vspace{0.4in}
\begin{center}
{\Large \bf  No Primordial Magnetic Field from Domain Walls\\}
\vspace{0.2in}
{\bf M.B. Voloshin  \\ }
Theoretical Physics Institute, University of Minnesota, Minneapolis, MN
55455 \\ and \\
Institute of Theoretical and Experimental Physics, Moscow, 117259
\\[0.2in]

{\bf   Abstract  \\ }
\end{center}

It is pointed out that, contrary to some claims in the literature, the
domain walls cannot be a source of a correlated at large scales
primordial magnetic field, even if the fermionic modes bound on the wall
had ferromagnetic properties. In a particular model with massive (2+1)
dimensional fermions bound to a domain wall, previously claimed to
exhibit a ferromagnetic behavior,  it is explicitly shown that the
fermionic system in fact has properties of a normal diamagnetic with the
susceptibility vanishing at high temperature.

\end{titlepage}

The existence of magnetic field correlated at a galactic scale \cite{bk}
is believed to require a strong primordial field correlated at
cosmological distances at some stage in the early universe. This
phenomenon would find a natural explanation, if the primordial field was
created by extended objects, having a cosmological size. One class of
extended objects that might have existed in the early universe is
provided by domain walls, assuming that a model containing these walls
successfully avoids general constraints \cite{zko} on undesirable
cosmological consequences of such extended defects. There has been
several claims in the literature \cite{iwazaki,ct1,ct2,fz} that the
modes of fermion field bound to a domain wall produce, in certain
models, a magnetic field $B$, which could provide a much needed
explanation of a primordial magnetic field in the early universe. The
purpose of this note is to point out that, as enticing as this
explanation could be, it cannot be physically correct. Namely, it is
almost trivial to show that, irrespective of the details of the
dynamics, a correlated over the entire area of a flat domain wall
magnetic field should be greatly suppressed by an inverse power of a
cosmological size, even if the wall exhibited a ferromagnetic behavior.
Furthermore, in the models that are sufficiently well formulated
\cite{iwazaki,ct1}, the walls are in fact diamagnetic, contrary to the
previous claims.  In this letter the model of Ref.\cite{ct1,ct2} with
massive fermionic modes with broken parity is considered, and it is
shown that the fermion system on the wall is  a normal diamagnetic,
while the previous claim of ferromagnetism is based on an incorrect
formula \cite{cea} for the $B$ dependent free energy of the fermionic
vacuum. \footnote{The claim of \cite{iwazaki} was explicitly shown
\cite{mv} to be based on an incorrect calculation of the energy of the
gas of massless (2+1) dimensional fermions in magnetic field.}

In order to prove the statement about the suppression of the magnetic
field that a domain wall might spontaneously create, let us introduce a
large normalization box with the sides of length $L$, and consider a
flat domain wall, spanning the box along the $x, \, y$ plane, thus $z$
being the coordinate axis perpendicular to the wall. From symmetry, a
correlated magnetic field can have only the $z$ component $B_z$, which
does not depend on the coordinates $x$ and $y$ parallel to the wall.
Then the Maxwell's equation div {\bf B} = 0 dictates that $B_z$ also
cannot depend on the coordinate $z$, i.e. $B_z$ has to be constant:
$B_z=B$. The energy of such constant magnetic field is proportional to
the volume $L^3$ of the box, while, the energy, associated with dynamics
on the wall is proportional to the area $L^2$ of the wall, so that the
total energy of the system is written as
\beq
E(B)=L^3 \, {B^2 \over 2} + L^2 \, f(B) ~~~
\label{engen}
\eeq
with $f(B)$ being the surface energy density in the presence of the
magnetic field, determined by specific dynamics on the wall. Clearly,
for any finite function $f(B)$ the value of $B$ providing the minimum to
the energy (\ref{engen}) goes to zero when the size $L$ of the
normalization box is taken to infinity. Thus one concludes that the
magnetic field of an infinite flat domain wall has to vanish,
independently of the specifics of the model.

The analyses of the papers \cite{iwazaki,ct1,ct2,fz} find a
ferromagnetic behavior for the function $f(B)$ at small field: $f(B)-
f(0) = -a \, B$, with the spontaneous magnetization $a$ being determined
by the `microscopic' parameters of the wall. According to
Eq.(\ref{engen})the generated field then is $B=a/L$.\footnote{Another
simple way to arrive at the same conclusion is to integrate over the
surface of the wall the magnetic field, created by each element of the
surface with constant density of magnetic dipole moment: the integral is
non-zero only due to edge effects, and vanishes, when the size of the
surface goes to infinity.} This consequence of Eq.(\ref{engen}) is
acknowledged only in the final version of Ref.\cite{iwazaki}, and is
ignored in Refs.\cite{ct1,ct2,fz}. In the latter papers the extent of
the field away from the wall is assumed to be of the order of the
thickness of the wall in direct contradiction with the Maxwell's
equation. Physically, a finite size $L$ can arise either as the Hubble
size \cite{iwazaki}, or as a domain size in a multi-domain structure. In
either case the magnitude of the correlated field is greatly suppressed
by $L^{-1}$ and is unlikely to be sufficient for explaining the required
primordial field. As an example, one can consider the estimates of the
field in the mechanism of Ref.\cite{fz}, where the primordial field is
associated with the axion domain walls at the QCD phase transition, i.e.
at the temperature $T_{QCD} \sim \Lambda_{QCD} \sim 0.2 \, GeV$. The
size of the correlations in the picture presented there is the Hubble
scale $l(T_{QCD}) \approx 30 \, km$. Thus even if one adopts the
estimate of Ref.\cite{fz} for the spontaneous magnetization: $a \sim e
\Lambda_{QCD}^2$, the estimated magnitude of the generated field should
be $B \sim e \, \Lambda_{QCD} /l(T_{QCD}) \sim 10^{-2} \, G$ (instead of
the claimed $10^{17} \, G$). Furthermore, it will be shown below that
the magnetization of the axion wall is proportional to the total baryon
number density accumulated on the wall, so that the estimate \cite{fz}
of the magnetization is rather on the maximalist side.

The existence of even a suppressed by $L^{-1}$ magnetic field is
contingent on ferromagnetism of the fermion modes on the wall. Such
behavior was claimed in Refs.\cite{cea,ct1,ct2} within a model of
massive (2+1) dimensional fermions. Moreover, it has been argued
\cite{ct2} that the ferromagnetic term $-B$ survives in the free energy
at all temperatures, in contrast with the known behavior in any other
systems. In what follows a calculation of the free energy of a massive
(2+1) dimensional fermion field is presented, in close analogy with a
similar calculation \cite{mv} for a massless case, and it is shown that,
contrary to previous claims, the system in fact exhibits a normal
diamagnetism with the diamagnetic susceptibility vanishing at high
temperature. The previous erroneous findings of the ferromagnetic
behavior were due to an inaccurate manipulation with a divergent sum in
Ref.\cite{cea}.

The model of massive (2+1) dimensional fermions is relevant to the
situation where the fermions have a non-zero mode bound on the wall with
eigenvalue $m$, and there is no mode corresponding to the eigenvalue
$-m$. (This un-pairing of the modes arises in a situation where the
P parity is broken.) The two-dimensional motion of charged fermions in
this mode is described by a (2+1) dimensional Dirac equation
\beq
\left ( i \, \gamma^\mu \, (\partial_\mu - i \, e \, A_\mu) -m \right )
\, \psi =0~,
\label{de}
\eeq
where $e$ is the electric charge, $A_\mu=(A_0, \, A_1, \, A_2)$ is the
vector potential of an (external) electromagnetic field, and
$\gamma_\mu$ is the set of $2 \times 2$ gamma matrices, which can be
chosen, e.g. in terms of the Pauli matrices as $\gamma^0=\sigma_3$,
$\gamma^1=i \, \sigma_1$, and $\gamma^2=i \, \sigma_2$. It is known (see
e.g. in Ref.\cite{cea}) that in an external magnetic field $B$,
perpendicular to the wall, there is an asymmetry between positive and
negative Landau energy levels, depending on the relative sign of $eB$
and $m$. Namely, assuming for definiteness that $m$ is negative and
$eB$ is positive, the spectrum of positive energy levels is given by
$E_{n+}=\sqrt{2 \, e \, B \, n +m^2}$ with $n=1,\, 2,\ldots$, while that
of the negative levels is $E_{n-}=-\sqrt{2 \, e \, B \, n +m^2}$ with
$n=0,\,1,\,2,\ldots$. In other words, the level with $n=0$ is absent
from the positive energy part of the spectrum, and is present in the
negative part.

The free energy per unit area for the fermion system at a temperature
$T=1/\beta$ is expressed through the spectrum in the standard way:
\beq
F=F_-+F_++E_{\rm vac}~,
\label{ffull}
\eeq
where
\beq
F_+=-\beta^{-1} {e \, B \over 2 \pi} \, \sum_{n=1}^\infty \ln \left( 1+
e^{-\beta \, \sqrt{2 \, e \, B \, n +m^2}} \right )
\label{fp}
\eeq
and
\beq
F_-=-\beta^{-1} {e \, B \over 2 \pi} \, \sum_{n=0}^\infty \ln \left( 1+
e^{-\beta \, \sqrt{2 \, e \, B \, n +m^2}} \right )
\label{fm}
\eeq
are the thermal parts of the free energy associated with the real gas of
fermions and antifermions at the Landau levels. The term $E_{\rm
vac}$ in Eq.(\ref{ffull}) is the energy of the vacuum state, i.e. with
all negative energy levels filled and those with positive energy being
vacant,
\beq
E_{\rm vac}= - {e \, B \over 2 \pi} \,\sum_{n=0}^\infty \sqrt{2 \, e \,
B \, n +m^2}~.
\label{evac}
\eeq

The sums for the temperature dependent parts $F_\pm$ are finite, and do
not cause controversy. The most interesting is the sum in
Eq.(\ref{evac}), which is divergent and thus should be handled with some
care. In order to make the latter sum tractable without ambiguity, it
needs to be regularized.  A gauge invariant regulator factor should
depend on a gauge invariant quantity, which naturally can be chosen as
the energy of the levels. The exact form of the regulator is a matter of
convenience, and we use here an exponential form of this factor:
$\exp(-\epsilon \, E_n^2)$, thus making the sum for the regularized
vacuum energy read as
\beq
E_{\rm vac}^{(r)}(B)= - {e \, B \over 2 \pi} \,\sum_{n=0}^\infty \sqrt{2
\,
e \, B \, n +m^2} \, \exp(-\epsilon \, 2 \, e \, B \, n - \epsilon \,
m^2)~,
\label{sr}
\eeq
where $\epsilon$ is the regulator parameter. In the physically
meaningful
quantity $E_{\rm vac}(B)-E_{\rm vac}(0)$, one should take $\epsilon \to
0$ in the final result.

The sum in Eq.(\ref{sr}) can be evaluated using Poisson's method based
on the identity:
\beq
\sum_{n=0}^\infty \, f(n) = \int_{\delta}^\infty \, f(x) \,
\sum_{n=-\infty}^\infty \, \delta(x-n) \, dx = \sum_{k=-\infty}^\infty
\,
\int_{\delta}^\infty \, f(x) \, \exp( 2 \, \pi \, i \, k \, x) \, dx~~,
\label{id}
\eeq
where $\delta$ is an arbitrary number, such that $-1 < \delta <0$.
Notice that the sum over $n$ in the second expression goes from
$-\infty$ to $+\infty$. The identity is still valid since the terms with
$n < 0$ are identically zero. The summand in Eq.(\ref{sr}) is
non singular at $n=0$, so that one can in fact set $\delta=0$, and write
\beq
E_{\rm vac}^{(r)}= - {e \, B \over 2 \pi} \,\sum_{k= -\infty}^\infty
\int_0^{\infty} \sqrt{2 \, e \, B \, x +m^2} \, \exp (-\epsilon \, 2 \,
e \, B \, x - \epsilon \, m^2 + 2 \, \pi \, i \, k \, x) \, dx~.
\label{srp}
\eeq
In this expression the only term in the sum over $k$ that is singular in
the limit $\epsilon \to 0$ is the one with $k=0$. This term  is given by
\beq
-{e \, B \over 2 \pi}\,
\int_0^{\infty} \sqrt{2 \, e \, B \, x +m^2} \, \exp (-\epsilon \, 2 \,
e \, B \, x - \epsilon \, m^2) \, dx = -{1 \over 4\pi} \,
\int_0^{\infty} \sqrt{z +m^2} \, \exp (-\epsilon \,z - \epsilon \, m^2)
\, dz
\label{k0}
\eeq
and does not depend on $B$. In fact this term is identically equal to
the vacuum energy, regularized in the same way:
\beq
E_{\rm vac}(0)=-\int \sqrt{p^2+m^2} \, \exp(-\epsilon \, p^2 - \epsilon
\, m^2) \, {d^2p \over (2\pi)^2}~.
\label{e0}
\eeq
Thus the $k=0$ term in the sum in Eq.(\ref{srp})
totally cancels in the difference $E_{\rm vac}(B)-E_{\rm vac}(0)$, and
the difference itself is given by the sum of all terms with $k \neq 0$.
The latter sum is finite in the limit $\epsilon \to 0$. However an
infinitesimal parameter $\epsilon$ should be retained in calculation of
the integrals in order to ensure convergence and proper phase definition
of the oscillatory integrals.  The sum over $k \neq 0$ is analytically
tractable in two limiting cases: large field, $e B \gg m^2$, and small
field, $e B \ll m^2$. In the large field limit the mass can be
neglected, and the result is given by the massless case \cite{cea,mv}:
\beq
E_{\rm vac}(B)-E_{\rm vac}(0)={ \zeta (3/2) \over 16 \, \pi^2} \, (2 \,
e \,
B)^{3/2}~,
\label{lb}
\eeq
corresponding to diamagnetism with a singular diamagnetic
susceptibility.

The most interesting here is the case of small field, since this is the
limit where the difference $E_{\rm vac}(B)-E_{\rm vac}(0)$ is claimed
\cite{cea,ct1,ct2} to have a linear dependence on $B$. Using the
representation in Eq.(\ref{srp}) we find however quite different result.
Namely, using Taylor expansion in $(2 \, e \, B /m^2)$ for the square
root in Eq.(\ref{srp}), and grouping together terms with $+k$ and $-k$,
one finds in the limit $\epsilon \to 0 \,$\footnote{Rearranging the sums
is justified as long as the series is absolutely convergent, i.e. at
finite $\epsilon$. The limit $\epsilon \to 0$ is taken after the
rearrangement.}:
\beq
E_{\rm vac}(B)-E_{\rm vac}(0)=-{e \, B \, |m| \over 2 \pi} \,
\sum_{s=0}^\infty \sum_{k=1}^\infty
{\Gamma({3\over 2}) \over \Gamma ({3 \over 2} -s)}
\left [(2 \, \pi \, i \, k)^{-s-1} + (-2 \, \pi \, i \, k)^{-s-1}
\right ]  \, \left( {2 \, e \, B \over m^2} \right )^s~.
\eeq
One can easily see that in the sum over $s$ only the terms with odd $s$
are
non vanishing, thus leaving only the even powers of $B$ in the
expansion. Denoting in the non-zero terms $s=2p+1$, one finally finds
the asymptotic expansion in powers of the field:
\beq
E_{\rm vac}(B)-E_{\rm vac}(0)={e^2 \, B^2 \over 2 \, \pi^3 \, |m|}
\sum_{p=0}^\infty {(-1)^p \, \zeta(2 \, p +2) \, \Gamma({3\over 2})
\over \Gamma ({1 \over 2} - 2 \, p)} \, \left ( {e \, B \over \pi \,
m^2} \right )^{2p}= {e^2 \, B^2 \over 24 \, \pi \, |m|} + \ldots ~.
\label{res0}
\eeq

The leading at small $B$ term in the expansion (\ref{res0}) is positive
and quadratic in $B$, thus corresponding to a normal diamagnetism. It is
satisfying to verify that, in compliance with the common physical
intuition, the diamagnetism vanishes at high temperature. By ``high" one
should understand a temperature constrained by the condition $T \gg m$,
since at lower temperatures the gas of real fermions cannot produce
substantial effect because of the thermal blocking factor $e^{-\beta
|m|}$. Notice, however that the temperature has to be still lower than
the
mass gap for the fermions on the wall, since otherwise the problem would
not be reduced to dynamics of a (2+1) dimensional fermion system, but
rather one would have to consider the full (3+1) dimensional problem. In
the latter situation however the effects of the wall are small, and no
significant phenomena are to be expected.

In order to calculate the thermal effects under the condition $T \gg m
\gg e \, B$ one can apply the Poisson summation formula (\ref{id}) for
the sums $F_+$ and $F_-$ in Eqs.(\ref{fp}) and (\ref{fm}). Writing
separately the term with $k=0$ and grouping together the terms with
symmetric non-zero values of $k$ results in the expression
\begin{eqnarray}
\label{fpr}
&&F_++F_-=-{e \, B \over \pi \, \beta}
\int_{\delta_1}^\infty \ln \left [ 1+ \exp(-\beta \, \sqrt{2 \, e \, B
\, x
+m^2}) \right ] \, dx- \\ \nonumber
&&{e \, B \over \pi \, \beta} \,
\left \{ {1 \over 2} \ln \left ( 1 + e^{-\beta |m|} \right ) +
\sum_{k=1}^\infty \int_{\delta_1}^\infty \ln \left [ 1+ e^{-\beta \,
\sqrt{2
\, e \, B \, x +m^2}} \right ] \, \left ( e^{2 \, \pi \, i \, k \, x} +
e^{-2 \, \pi \, i \, k \, x} \right ) \, dx \right \} ~.
\end{eqnarray}
Here the lower limit of integration, $\delta_1$, is such that $0 <
\delta_1
< 1$, corresponding to summation over the Landau levels from $n=1$ to
$\infty$. (The contribution of the $n=0$ level in the negative energy
spectrum is the first term in parenthesis.) The first term on the right
hand side of Eq.(\ref{fpr}) arises from the $k=0$ harmonic in the
Poisson formula. Taking in this term the limit $\delta_1 \to +0$, one
sees that this term does not depend on $B$ and describes the
thermal part of the free energy of the free fermion gas at zero field:
\beq
F_0(\beta, \, m)= -2 \beta^{-1} \int \ln \left [ 1+
\exp\left(-\beta \, \sqrt{p^2 +m^2}\right ) \right ] \, {d^2p \over
(2\pi)^2}~.
\label{f0b}
\eeq
The expression in the parenthesis in
Eq.(\ref{fpr}) vanishes at $B=0$, since the first term cancels against
the sum due to the identity
\beq
\sum_{k=1}^\infty \, \int_{\delta_1}^\infty
\, \left ( e^{2 \, \pi \, i \, k \, x} + e^{-2 \, \pi \, i \, k \, x}
\right ) \, dx = -{1 \over 2}~.
\label{i0}
\eeq
The rest of the terms
can be found by Taylor expansion in $B$ of the integrand in the
parenthesis. For the first non vanishing term one finds
\begin{eqnarray}
&&F_++F_--F_0(\beta, \, m)=-{e \, B \over \pi \, \beta} \left \{ -
{\beta \, e^{-\beta |m|} \over (1+e^{-\beta |m|})} \, {e \, B \over |m|}
\, \sum_{k=1}^\infty \int_{\delta_1}^\infty x \, \left ( e^{2 \, \pi \,
i \,
k \, x} + e^{-2 \, \pi \, i \, k \, x} \right ) \, dx \right \}
\nonumber \\
&&+ O(B^4)=- {2 \,  e^{-\beta |m|} \over 1+e^{-\beta |m|}}
\, {e^2 \, B^2 \over 24 \, \pi \, |m|}+ O(B^4)~.
\label{ff}
\end{eqnarray}
Combining this result with Eq.(\ref{res0}), one readily
sees that the diamagnetic susceptibility indeed vanishes at $\beta |m|
\ll 1$, i.e. at $T \gg |m|$.

The presented calculation of the free energy of the fermionic system
leads us to the conclusion that in the considered model the fermionic
mode bound to the domain wall gives rise to a normal diamagnetism. As
usually, the diamagnetic behavior for a system of charged fermions
arises because the diamagnetism of the `orbital' motion of the electric
charges in the magnetic field overcomes the paramagnetism, associated
with the normal magnetic moment. In order to avoid the diamagnetism, it
was suggested \cite{fz} that neutral fermions with an anomalous magnetic
or electric dipole moment, bound to a domain wall, are fully polarized
and provide a spontaneous magnetization. Specifically, the strongest
effect of the model of Ref.\cite{fz} arises from a mode for neutrons,
bound to an axion domain wall right after the QCD chiral transition.
Here we present some remarks on this model.

In terms of a (2+1) dynamics of the bound fermions the Hamiltonian for
the anomalous interaction has the form
\beq
H_{\rm an}=-\kappa {\overline \psi} \sigma^{\mu \nu} \, F_{\mu \nu} \,
\psi~,
\label{han}
\eeq
where $\kappa$ is the anomalous magnetic moment, and $\sigma^{\mu
\nu}={i \over 2} (\gamma^\mu \, \gamma^\nu - \gamma^\nu \, \gamma^\mu)$.
For a magnetic field $B$ in the $z$ direction, $B=F_{12}$, one has
$\sigma^{12}=\gamma^0$, so that the Hamiltonian takes the form $H_{\rm
an}=-\kappa \, B \, \psi^\dagger \psi$. Thus the spontaneous
magnetization $a=-(\partial F/\partial B)|_{B=0}$ coincides, up to the
factor $\kappa$ with the density of the fermionic charge: \beq a=\kappa
\, \langle \psi^\dagger \psi \rangle~, \eeq where the averaging over the
appropriate thermal state is implied. For the model of Ref.\cite{fz} the
net surface density of neutrons $\nu_B=\langle \psi^\dagger \psi
\rangle$, occupying the zero mode on the axion wall, is limited by at
least two factors. One is the maximal occupation density for the mode
set by the requirement that the Fermi energy does not exceed the energy
gap $\Delta$ between the modes: $\nu_B ~^<_\sim ~\Delta^2$. The value of
the gap is model dependent, however, one can perhaps take $\Delta \sim
\Lambda_{QCD}$ for an estimate. The other limit arises from a
consideration of the diffusion of netrons from the bulk to the wall.
Because of low density of the baryon number in the bulk $n_B$ it takes
time for the netrons to be accumulated on the wall. Taking the mean free
path for the neutrons at $T_{QCD} \approx \Lambda_{QCD}$ as $ \sim
1/\Lambda_{QCD}$ one estimates the density of neutrons accumulated by
the wall over the time $t$ as $\nu_B \sim n_B \, \sqrt{t
/\Lambda_{QCD}}$. Under a maximalist assumption, that the available
diffusion time $t$ is given by the Hubble time, $t \sim l(T_{QCD})$, one
finds that at $n_B/T^3 \sim 10^{-10}$ this limit for $\nu_B$ {\it
numerically} is also close to $\Lambda_{QCD}^2$.

Summarizing the discussion of this paper, we conclude that the gas of
modes of charged Dirac fermions on a domain wall is always diamagnetic
with susceptibility vanishing at high temperature. Therefore such
systems cannot spontaneously produce any correlated magnetic field. The
spontaneous magnetization of the modes for neutral fermions with
anomalous magnetic moment is proportional to the surface density of the
fermion number on the wall, and  the magnetic field generated by a
spontaneous magnetization of a wall is suppressed by the inverse of the
wall's dimension. These considerations lead us to the general conclusion
that it is at least highly unlikely that domain walls can provide a
physically viable source of the primordial magnetic field.

This work is supported in part by DOE under the grant number
DE-FG02-94ER40823.


\begin{thebibliography}{99}
\bibitem{bk}
R. Beck {\it et. al.}, Ann. Rev Astron. Astrophys. {\bf 34} (1996) 155;
P. Kronberg, Rep. Prog. Phys. {\bf 57} (1994) 325.
\bibitem{zko}
Ya.B. Zeldovich, I.Yu. Kobzarev and L.B. Okun, \spj{40}{74}{1}.
\bibitem{iwazaki}
A. Iwazaki, \pl{B 406}{97}{304}; \prev{D 56}{97}{2435}.
\bibitem{ct1}
P. Cea and L. Tedesco, \pl{B 425}{98}{345}.
\bibitem{ct2}
P. Cea and L. Tedesco, \pl{B 450}{99}{61}.
\bibitem{fz}
M.M. Forbes and A.R. Zhitnitsky, Univ. of British Columbia report, May
2000;~~[hep-ph/0004051]
\bibitem{mv}
M.B. Voloshin, \pl{B 389}{96}{475}.
\bibitem{cea}
P. Cea, \prev{D 32}{85}{2785}; \prev{D 34}{86}{3229}.


\end{thebibliography}
\end{document}